# Third Harmonics of the AC Magnetic Susceptibility: a method for the study of flux dynamics in High Temperature Superconductors

Shortened title: "**Flux dynamics in HTS studied by the third harmonics of AC Susceptibility**"

M. Polichetti, M.G. Adesso, S. Pace

Dipartimento di Fisica, Universitá degli Studi di Salerno & INFM
Via S. Allende, Baronissi (Salerno), 84081, Italy
Phone: ++39 89 965 307; Fax: ++39 89 953 804;
e-mail: adesso@sa.infn.it

## Abstract

The temperature dependence of the 1$^{st}$ and 3$^{rd}$ harmonics ($\chi_{1,3}^{',''}$) of the AC magnetic susceptibility has been measured on melt grown YBCO samples for different frequencies and amplitudes of the AC magnetic field and intensity of a contemporaneously applied DC field. With the help of critical state models and of numerical simulations [22], we have devised a novel method, based on the combined analysis of the 1$^{st}$ and the 3$^{rd}$ harmonics (specifically on the comparison between $\chi_1^{''}$ and $\chi_3^{'}$), that allows to distinguish different temperature ranges dominated by the different dissipative magnetic flux regimes. In particular, we identified three principal "zones" in the temperature dependence of the real part of the 3$^{rd}$ harmonic: the "zone 1", in the temperature range below the peak of the imaginary part of the 1$^{st}$ harmonic, $T_p(\chi_1^{''})$, and the "zone 2", characterized by $\chi_3^{'}$ negative values in a temperature region just above $T_p(\chi_1^{''})$, both dominated by the creep regime; the zone 3, just below $T_c$, in which we revealed the presence of Thermally Assisted Flux Flow

**Corresponding author:** Maria Giuseppina Adesso, Department of Physics, University of Salerno - INFM - 84081 Baronissi (SA), Italy - Phone: ++39 89 965 307; Fax: ++39 89 953 804;
e-mail:adesso@*sa.infn.it*

(TAFF). By the identification of these "zones", an estimation of the value of the pinning potential can be obtained.

PACS Codes: 74.25.Ha, 74.60.Ge

KEYWORDS: vortex dynamics, harmonics of AC susceptibility, TAFF, pinning potential

# 1. INTRODUCTION

Magnetic properties of the High Temperature Superconductors (HTS) have been widely studied in order to understand the mechanisms governing the flux lines dynamics, and to find the key to improve the application properties of these materials. Although the same kind of information can be extracted by means of direct transport measurements, the high sensitivity of the AC susceptibility technique gives the opportunity to investigate the low voltage region in the I-V characteristics, such as the TAFF regime. These studies have shown that the magnetic response of a superconductor can be both linear and nonlinear, which is characterized by the presence of higher harmonics [1,2].

The meaning of the first harmonic is clear: its real part is associated to the screening properties of the sample, and the imaginary part is proportional to the energy converted into heat during one cycle of AC field [3,4]. On the other side, the physical interpretation of the higher harmonics of the AC susceptibility is still under discussion, as well as the strong differences in the shape of their curves measured in different conditions still need clarifications.

The existence of higher harmonics, due to hysteretic losses, was first predicted by C.P. Bean [5,6]. In his frequency independent critical state model, in which the critical current density ($J_c$) is independent of the magnetic field (*B*), the even harmonics are zero. Nevertheless, non zero even harmonics are predicted, by the Kim critical state model, in which a magnetic field dependent $J_c$ is considered [7-9]. Shatz et al. [10] showed that all the critical state models can be described in terms of a unique parameter δ, which measures the extent of penetration of the AC field into the slab. δ depends on the slope *S* of the field profile inside the sample (the slope *S* being a function of the temperature and the field dependent $J_c$), on the applied magnetic field $h_{ac}$ and on the geometrical dimension of the sample. In particular, for a slab with thickness 2*a*, δ=$h_{ac}$/(*S*•*a*). In the Bean model δ=$h_{ac}$/H*, where H* is the smallest field that penetrate the whole sample. The AC susceptibility harmonics, calculated by using any critical state model, are independent of the frequency ν of the applied AC field [10]. Nevertheless, many experimental results show that variations of all the harmonics $\chi_n$ appear in correspondence of different AC frequencies. In particular, the fundamental harmonic $\chi_1$ show a change in the width of the superconducting transition both in the real ($\chi_1'$) and in the imaginary component ($\chi_1''$), and a variation of the height and of the temperature $T_p(\chi_1'')$, corresponding to the peak position in $\chi_1''$, whereas the general shape of the curves of both the components remains unchanged [11, 12]. For the 3rd harmonic, different AC frequencies produce variations in the height and in the temperature of the peak maximum of the $|\chi_3|$ modulus and of its breadth [13,14]. The variation of the temperatures $T_p(\chi_1'')$ and $T_p(|\chi_3|)$ can be explained by modifying the critical state models, taking into account the presence of relaxation phenomena, i.e. by introducing a critical current density $J_c(\nu)$ which increases with the frequency $\nu$ [14,15]. However, these modified critical state models are not able to explain the variations of the $\chi_1''$ and $|\chi_3|$ peak heights and the large variations in the shape of the two separate components $\chi_3'$ and $\chi_3''$, which experimentally appear as the frequency is changed [16-18].

The study of these variations can be usefully tackled by using numerical techniques, which integrate the diffusion equation of the magnetic field inside the sample, where the diffusion processes are determined by the resistivity $\rho$. From the numerical solutions of the equation, the harmonics (in particular the 3rd) of the AC susceptibility have been computed both in presence and in absence of a DC external magnetic field by considering only the presence of creep phenomena ($\rho = \rho_{creep}$) [19,28], this giving also a good method for the estimation of the pinning potential [20,21].

The diffusion equation has been also solved by including all the dynamic regimes (Flux Creep, Flux Flow, Thermally Activated Flux Flow or TAFF, and the parallel of Flux Creep and Flow) without a DC magnetic field [22]. At fixed frequency (and more generally at fixed external parameters), the shape of the 3rd harmonics also shows significant variations, essentially due to different pinning properties [22, 23]. Moreover shape variations due to the sample geometry have been predicted [24-26].

Nevertheless, up to now, a detailed comparison between experimental results and numerical simulations is missing.

The aim of this work is the analysis of the temperature dependence of the 3rd harmonics measured on melt grown YBCO samples, in order to determine the connection between the detailed structure in the shape of the curves and the flux dynamic regimes. This has been obtained by measuring the AC susceptibility on different samples with the same geometry, thus excluding all the variables related to the sample geometrical characteristics. In particular we will illustrate a new method, based on the comparison of both the curves of the 1st and 3rd harmonics, allowing us to detect the flux regimes which are predominant in the different temperature ranges of the $\chi_{1,3}^{',''}$ curves.

The method does not depend on the analyzed sample, and it can be straightforwardly adapted to the analysis of the AC magnetic response of other bulk high-$T_c$ superconductor samples with similar geometric properties. So, although we tested the method on different samples, for simplicity here we only reported the data referred to one sample.

## 2. EXPERIMENTAL DETAILS

An in-house made AC susceptometer has been used to measure the temperature dependence of the first ($\chi_1^{',"}$) and the third ($\chi_3^{',"}$) harmonics of the AC susceptibility on different melt grown $Y_1Ba_2Cu_3O_{7-\delta}$ samples, obtained by cutting a 10 cm long directionally solidified bar (fabricated following the procedure described in Ref. [27]) from the central part of the sample, where the homogeneity was checked to be the highest. In this work we will refer mainly to a homogeneous slab with dimensions 1.8 mm x 3.2 mm x 5 mm, cut exactly at the geometrical center of the bar, although similar results have been found on the other samples.

First characterization of the sample was performed by means of magnetization curves, and the values $J_c$ (77K; $H_{DC}=0$) = $8.4 \times 10^7 A/m^2$ (corresponding to $J_c(0K; H_{DC}=0) \approx 10^9$ $A/m^2$) and $T_c$(onset)$\approx$91.6 K were determined.

In susceptibility measurements, particular care has been taken to maximize the thermal contact between the thermometer and the sample, and a very slow temperature sweep (0.1K/min) has been used in order to optimize the thermal stability, to minimize the error on the temperature readings and to avoid thermal gradients along the sample.

Moreover, the sensitivity of the 3$^{rd}$ harmonic curve shape to the phase setting of the measurements forced us to choose it in a very accurate way: for every frequency, the phase has been set such that the imaginary part of the first harmonic is zero, at the lowest AC field ($h_{AC}$ = 0.5 Oe) and at a temperature T<<T$_c$ (T≈52 K). We also verified that the same phase results from a setting done at T>T$_c$, where dissipations are negligible, being the skin depth $\delta_n$ much larger than the characteristic size of the sample (for example $\delta_n$=8 cm for ν=107 Hz). In this way, the phase uncertainty was lower than 0.1 degree.

Measurements were performed applying magnetic fields parallel to the longitudinal axis of the sample, at different AC field amplitudes ($h_{AC}$ = 1, 2, 4, 6, 8, 16, 32 Oe) and frequencies ($\nu$ = 10.7, 107, 1607, 10700 Hz), both with and without an external DC field (H$_{DC}$ = 0, 100, 200, 400 Oe).

## 3. EXPERIMENTAL RESULTS AND DISCUSSION

### 3.1 *AC susceptibility measurements with H$_{DC}$ = 0*

In Fig. 1*a*) and in the Inset, the temperature dependences of the real component of the first harmonic $\chi_1'(T)$ are plotted for different $\nu$ and $h_{AC}$ respectively, with $H_{DC}$=0. The dependence of $\chi_1(T)$ on the AC field amplitude confirms a non-linear dynamic response. As the AC amplitude increases, a general broadening of the transition appears, the step-like transition in the real part $\chi_1'(T)$ becomes always smoother, whereas in the corresponding imaginary part (not shown) a decrease of the temperature of the peak ($T_p$), an increase of the peak height, $(\chi_1'')_{MAX}$, towards the value 0.24 (predicted in the Bean critical state model [5,6]), and a general widening of the peak appear. Nevertheless, such broadening does not give us any detailed information about the flux dynamic regime governing the magnetic response of the sample. On the other hand, the measurements show that the first harmonic of the AC susceptibility is just weakly dependent on the frequency of the AC field, as reported in Fig. 1 for the real component. This frequency independent $\chi_1(T)$, together with its dependence on the AC field amplitude, suggests that the detected magnetic response can be interpreted mainly in terms of a critical state model, with a negligible contribution coming from dynamic effects.

On the contrary, the analysis of the 3rd harmonics shows that the curves depend strongly also on the AC field frequency, as reported in Fig. 1*b*), and not only on its amplitude (see inset of Fig. 1*b*)). For this reason, the third harmonic represents a very sensitive tool to investigate the flux dynamics.

As it has been reported in a previously published paper [18], the $\chi_3$(T) measurements on an YBCO sample coming from the same batch are in qualitative agreement with the results obtained by numerical simulations [22] of diffusion processes in the framework of the collective pinning model [29, 30]. Therefore, in the present work, as far as the comparison with numerical simulations is concerned, only this model will be considered for the $U_p$ and $J_c$ temperature dependences.

A combined analysis of the simulated behavior of the 1st and 3rd harmonics of the AC susceptibility, together with the results predicted by the Bean critical state model suggest that it is possible to identify the different vortex dynamics regimes governing the AC magnetic response of the sample.

Indeed, in the Bean model the value $\chi_3'(T)$ is zero if the parameter δ<1 (i.e. if H<H*). For a cylindrical sample, the temperature at which the AC field amplitude is equal to the full penetration field (δ=1) is the peak temperature ($T_p$) in $\chi_1''$, i.e. T(δ=1)=$T_p$($\chi_1''$)≡$T_p$ [2,31]. For a slab, this equality is valid within an error of ~1%. Therefore, in the Bean critical state model, $\chi_3' = 0$ for $T<T_p$. Moreover, since in the critical state models the $\chi_3'$ can only assume values ≥ 0, the presence of negative values of $\chi_3'$ and $\chi_3'\neq 0$ values for $T<T_p$ indicates that the sample response is governed by dynamic phenomena which can not be described in terms of the Bean model. From the diffusion equation simulations [19, 22, 28], it is possible to identify the different dynamical regimes that are experimentally evident both for $T>T_p$ and for $T<T_p$.

In Figure 2 the experimental $\chi'_3(T)$ curve, taken at $\nu=1607$ Hz, $h_{AC}=8$ Oe, and $H_{DC}=0$, is shown, where no-zero values are present for $T<T_p$ (let's define this region: "zone 1") and negative values can be found in two different portions for $T>T_p$: the "zone 2" at temperatures slightly higher than $T_p$, between $T_p$ and a temperature we call $T^*_1$, and the "zone 3" at temperatures slightly lower than the onset of $\chi'_3$, $T_{3on}(\chi'_3)$ between a temperature indicated with $T^*_2$ and $T_{3on}(\chi'_3)$.

Experimentally, these zones have been found to have different frequency dependences, suggesting that they can be due to different dynamical regimes.

By analyzing the $\chi'_3(T)$ simulated curves [22], obtained by considering the TAFF resistivity in the diffusion equation ($\rho=\rho_{TAFF}$) [Eq. 2 in 22], the influence of the TAFF can be definitely neglected for temperatures well below $T_p$, where give a negligible contribution. On the contrary, the $\chi'_3(T)$ curves, simulated by imposing the presence of the Flux Creep (i.e. $\rho=\rho_{Creep}$ in the diffusion equation) [19,22] suggest that the main contribution to the experimental magnetic behavior in both the "zone 1" and the "zone 2" can be ascribed to creep phenomena. However, in principle the contribution of flux flow in these zones cannot be completely neglected. On the other side, the "zone 3" can not be described in terms of flux creep events, since, in proximity of $T_{3on}(\chi'_3)$, the $\chi'_3(T)$ curves simulated with $\rho=\rho_{Creep}$ assume only positive values. Therefore, the dynamic regimes which can influence the "zone 3" can only be the Flux Flow and the TAFF.

It is worth mentioning that the presence of all the different zones is not identified in all the performed $\chi'_3(T)$ measurements, since they are dependent on the experimental parameters $\nu$, $h_{AC}$ and $H_{DC}$, which determine the dynamic regime governing the AC magnetic response of the sample. For example, in the measurements at $\nu =1607$ Hz the existence of the "zone 2" is always evident for all the applied AC field amplitudes, and the size of this zone increases for increasing $h_{AC}$, as shown in the Figures 3a) and 3b) for $h_{AC} = 1$ Oe and $h_{AC} = 16$ Oe respectively. On the other hand, at $\nu=107$ Hz the "zone 2" is not present for fields $h_{AC}$ up to 8 Oe (see Figure 4a)), but at higher AC fields it becomes dominant and coalesces with the "zone 3", as reported in Figure 4b).
Nevertheless, some characteristics common to all the measurements at $H_{DC}=0$ can be summarized:

- the "zone 1" is always present for any $\nu$ and $h_{AC}$,
- at fixed $h_{AC}$ and for increasing frequencies, the minimum in $\chi'_3(T)$

at $T<T_P$ becomes deeper,

- as far as the AC frequency increases a reduction of the height of the $\chi_3'(T)$ maximum occurs in the region where the curves are similar to the Bean model ones,

- as the AC field increases, at fixed $\nu$, the "zone 2" grows towards higher temperature and the minimum in the "zone 1" becomes more pronunced.

These behaviors suggest an enhancement of the dynamic effects, when $\nu$ and $h_{ac}$ increase, very likely due to the flux creep regime.

### *3.2 AC susceptibility measurements with $H_{DC} \neq 0$*

The same analysis has been performed also in presence of a DC field ($H_{DC}$) up to 400 Oe.

All the 1$^{st}$ harmonics measured with $H_{DC}\neq 0$ (not reported here) show some common features. First of all, for increasing $H_{DC}$, the width of the transition (and, then, the width of the peak in $\chi_1''(T)$) increases, especially at higher AC field amplitudes. Moreover, both the temperature $T_P$ and the temperature of the onset in the $\chi_1'(T)$ curve moves towards lower temperatures, according to a critical state description, although the value of $(\chi_1'')_{MAX}$ is always smaller than the value predicted in the critical state models. However, the value $(\chi_1'')_{MAX}$ increases with the AC field amplitude (at fixed $H_{DC}$), being more evident at low $h_{AC}$ and $H_{DC}$. In general, the effect of an increasing $h_{AC}$ on the 1$^{st}$ harmonics is similar to what is produced by an increase of $H_{DC}$, and the general behavior is in agreement with both the measured and simulated data reported in literature [11, 32]. Finally, the presence of a non-zero DC field amplifies the decrease of $T_P$ and the increase of $(\chi_1'')_{MAX}$ produced by the use of higher frequencies $\nu$, which cannot be described in terms of critical state [11,12,28]. Also in the case with $H_{DC} \neq 0$, the information about the flux dynamics regimes can be extracted by using the combined analysis of the 1$^{st}$ and the 3$^{rd}$ harmonics.

In Figure 5 the typically measured temperature dependences of $\chi_1''$ and $\chi_3'$ obtained with $H_{DC} \neq 0$ are plotted, as representative of all the measurements in presence of a DC field, to illustrate this procedure. Also in the presence of a DC field, we can identify different "zones" dominated by non-linear dynamic phenomena, but these regions are different from the previous case. In particular, for $T \leq T_{3on}(\chi_3'')$ the "zone 3" with $\chi_3' < 0$, previously identified in the measurements with $H_{DC}=0$, is not present anymore. The variation with the DC field of the $\chi_3'$ temperature dependences is reported in Figure 6.a, where the difference due to the application of a DC magnetic field is magnified in the inset. Analogously, in Fig. 6.b), at $H_{DC} \neq 0$, close to $T_{3on}(\chi_3'')$, the $\chi_3''(T)$ curves exhibits a negative minimum, as it is expected [6] from critical state models, which is absent at $H_{DC}=0$ (see the inset of Fig. 6.b)). These behaviors were expected from the indications given in the previous analysis of the curves at $H_{DC}=0$: the presence of the "zone 3" is to be attributed to the TAFF or the Flux Flow regimes. In presence of magnetic fields $H_{DC}>>h_{AC}$, the total field is almost constant, so that the field dependence of the Flux Flow and of the TAFF resistance disappears, generating a linear diffusion process of the magnetic field, so that there is no higher harmonic contribution.

On the other hand, the previously identified "zone 2" is still present for any DC field, but only for low AC field amplitudes. In fact, for $\nu=1607 Hz$ the "zone 2" disappears for increasing $h_{AC} \geq 4$ Oe, as it is shown in Figure 7 by the comparison between the $\chi_3^{''}(T)$ curves, taken at various $h_{AC}$, and the relative $T_p$ extracted from the corresponding $\chi_1^{''}(T)$ measurements.

In any case, for $T<T_p$, the $\chi_3^{'} \neq 0$ values prove the presence of dynamic phenomena which are determined by the flux creep, that is the only dynamic regime still non linear even in presence of a static magnetic field $H_{DC}>> h_{AC}$.

This statement is confirmed by the behaviour of the $\chi_3^{'}(T)$ curves measured at different frequencies (see inset of Figure 7). In fact, the increase of the AC frequency increases the contributions due to flux dynamics processes that modify the behaviour due to the critical state. This result is in agreement with the simulations [22] and the measurements obtained with $H_{DC}=0$.

### 3.3 The TAFF regime and the estimation of the pinning potential

In the previously described method, the comparison between the simulated and the measured curves of the AC harmonic susceptibility allowed us to detect the presence of the different flux dynamic regimes which dominate the experimental AC susceptibility curves in the different temperature ranges. In particular, by comparing the 3$^{rd}$ harmonic in absence and in presence of $H_{DC}$ fields respectively, the existence of the zone 3, for $T \leq T_c$, governed by TAFF or Flux Flow resistivities, has been found.

The numerical simulations in presence of a DC magnetic field [32] show that a characteristic temperature $T^*$ exists, above which the $\chi_1^{''}(T)$ behavior is governed by linear TAFF resistance, since the Flow resistance is high that it gives a negligible contribution to the first harmonic. This $T^*$ is nearly coincident with the onset temperature $T_{3on}(\chi_3^{''})$, i.e. the temperature below which the imaginary part of the third harmonic $\chi_3^{''}$ starts to assume non zero values. Therefore, a difference $\Delta T_{on}$ between $T_{3on}(\chi_3'')$ and $T_{on}(\chi_1'')$ represents the experimental evidence of the presence of linear TAFF phenomena in the system under analysis.

This difference has been revealed in the detailed measurements of $\chi_1''(T)$ and $\chi_3''(T)$ performed in presence of an external $H_{DC}$>> $h_{AC}$ on our samples in the temperature range around the transition temperature, at different AC field amplitudes and frequencies.

In Figure 8 a clear example of the presence of the linear TAFF regime is reported for a particular set of experimental parameters: $H_{DC}$=400 Oe >> $h_{AC}$=1Oe. On the contrary, for the same sample, measured in the same conditions but with $H_{DC}$=0 (see inset of Fig.8), it is not possible to define unambiguously a value $\Delta T_{on}\neq 0$ within the sensitivity of our apparatus.

From the previous considerations, therefore, it is possible to state that the magnetic behavior of our samples can be divided in 3 main regions:

*i)* the region of $T<T_p$, where the flux creep regime is prevailing and then the pinning potential $U_p(T)$>>$k_B T$;

*ii)* an intermediate region with $T_p<T<T_{3on}(\chi_3'')$ where it is not easy to distinguish between the critical state and the flux creep contributions to the harmonic response;

*iii)* the region of $T_{3on}(\chi_3'')<T<T_{on}(\chi_1'')$ in which the magnetic response is dominated by the TAFF regime and then $U_p(T)<k_BT$.

Insofar it is reasonable to assume that, at the temperature $T=T_{3on}(\chi_3^")$, we have:

$$U_p(T_{3on}(\chi_3^"),B) \approx k_B T_{3on}(\chi_3^"). \quad (1)$$

The agreement [18] between our experimental data and the simulations obtained by considering the collective pinning regime, suggests us to use the collective pinning temperature dependence for the pinning potential [29, 30]; so that the Eq. (1) can be written as:

$$\frac{U_p(0,B)}{k_B} \approx \frac{T_{3on}(\chi_3^")}{\left(1-\left(\frac{T_{3on}(\chi_3^")}{T_c}\right)^4\right)} \quad (2)$$

where for $T_c$ the onset temperature of the first harmonic can be assumed. If we consider the measurements reported in Fig. 8, for $H_{DC}$=400 Oe, the values $T_c$=(91.17±0.01) K and $T_{3on}(\chi_3^")$=(90.99±0.01) K can be extracted. By substituting these values in the Eq. (2), and using [19, 32] the Kim-type dependence:

$$\frac{U_p(0,B)}{k_B} = \frac{U_p(0,0)}{k_B} \cdot \left(\frac{B_0}{B+B_0}\right) \quad (3)$$

with $B_0$=1T, $B=H_{DC}$, we can obtain the estimation of the pinning potential $U_p(0,0)$ at zero magnetic field and temperature:

$$\frac{U_p(0,0)}{k_B} = (1.1 \pm 0.1) \times 10^4 \text{ K}.$$

In order to verify the correctness of our previous assumptions, this value can now be used to calculate the potential $U_p(T,B)$ in the zone 1 and zone 3 where respectively the flux creep and the TAFF phenomena were found to be dominant, using the following formula:

$$\frac{U_p(T,B)}{k_B} = \frac{U_p(0,0)}{k_B}\left(1-\left(\frac{T}{T_c}\right)^4\right)\cdot\left(\frac{B_0}{B+B_0}\right) \qquad (4)$$

In fact, for $T=88$ K (zone 1) we obtain the value $\frac{U_p(88K,400G)}{k_B}=(1.4\pm 0.1)\times 10^3$ K $\gg$ 88K, and for $T=91.1$ K (zone 3) we obtain $\frac{U_p(91.1K,400G)}{k_B}=(34\pm 8)$ K $<$ 91.1 K. These results confirm the validity of our analysis in terms of the flux dynamic regimes acting in the different temperature regions of the AC susceptibility experimental curves.

## 4. CONCLUSIONS

In this work we have shown that the analysis of only the fundamental harmonics of the complex AC susceptibility is not sufficient to correctly interpret the magnetic response of YBCO samples in terms of flux dynamics. On the contrary, the higher sensitivity of the third harmonics to the variation of the experimental parameters and, therefore, of the magnetic non linear diffusion phenomena inside the sample, suggests to use these higher AC susceptibility harmonics to investigate the contribution of the different flux dynamic regimes to the magnetic response of the sample.

Therefore, starting from the general critical state model considerations, and from the recent numerical simulations of the AC magnetic response in presence of the different dissipative phenomena [22], a method, based on the combined analysis of the 1$^{st}$ and the 3$^{rd}$ AC harmonics, has been developed in order to identify the different flux dynamic regimes which are dominant in the various temperature ranges. Its application to the experimental results, obtained on YBCO melt textured samples, allowed us to identify three distinct "zones" in the $\chi'_3(T)$ measurements, depending on the experimental parameters $\nu$, $h_{AC}$ and $H_{DC}$, and their interpretation has been given in terms of defined dynamic regimes corresponding to each one.

In particular, since in the presence of a DC field $H_{DC} \gg h_{AC}$ the TAFF regime is linear, the "zone 1", individuated by no zero values of $\chi'_3(T)$ for T<$T_p(\chi''_1)$ ) and the "zone 2", where $\chi'_3(T)$ assumes negative values in a temperature range just above $T_p(\chi''_1)$ (namely $T_p(\chi''_1)$<T<$T^*_1$), are dominated by the flux creep regime.

On the contrary, the AC response of the sample in the "zone 3",

just below the transition temperature, is dominated by the TAFF. This last regime can be identified as the temperature range near $T_c$ where negative values of $\chi_3^{''}(T)$ disappear by applying a DC field $H_{DC} \gg h_{AC}$, while $\chi_1^{''}(T)$ is still present, resulting in a temperature difference between the onset of the 1$^{st}$ harmonic and of the 3$^{rd}$ one.

The qualitative agreement between the numerical and the experimental curves is good, and the quantitative differences can be attributed to the non optimized choice of the material parameters used for the simulation.

Nevertheless, a quantitative estimation of the pinning potential of our sample has been derived.

**ACKNOWLEDGEMENTS**

The authors are very grateful to Dr. A. Vecchione and M. Boffa for their YBCO samples, to Dr. E. Martinucci for the useful discussions, and to Mr. A. Ferrentino and Mr. F. Vicinanza for their technical support.

**Figure Captions**

**Fig. 1a)**: The weak frequency dependence of the real part of the 1$^{st}$ harmonics of the AC susceptibility $\chi_1'(T)$, plotted as function of the temperature for different frequencies $\nu$ and with fixed $h_{AC}$ and $H_{DC}=0$; **Inset of Fig. 1a)**: $\chi_1'(T)$ at different AC field amplitudes $h_{AC}$ and for fixed $\nu$ and $H_{DC}=0$; **Fig. 1b)**: The relatively strong frequency dependence of the 3$^{rd}$ harmonics $\chi_3'(T)$ corresponding to the $\chi_1'(T)$ curves in Fig. 1a): even the shape of the $\chi_3'(T)$ curves change with the frequency; **Inset of Fig. 1b)**: The $\chi_3'(T)$ curves measured at different AC field amplitudes $h_{AC}$ and for fixed $\nu$ and $H_{DC}=0$.

**Fig. 2**: An experimental $\chi_3'(T)$ curve, measured at $\nu=1607$ Hz, $h_{AC}=8$ Oe, and $H_{DC}=0$, summarizing the analysis discussed in the text for $H_{DC}=0$ (quadrants Q$_1$, Q$_2$, Q$_3$, and Q$_4$), and evidencing the existence of the 3 different individuated zones: the "zone 1" (vertical lines fillled), i.e. the $\chi_3'(T)$ no-zero values for T<$T_p(\chi_1'')$; the "zone 2" (horizontal lines filled), characterized by negative values of $\chi_3'(T)$ near $T_p(\chi_1'')$, restricted between $T_p(\chi_1'')$ and a temperature T*$_1$ (dashed line); the "zone 3" (crossed lines filled), with negative values just below T$_c$, between a temperature T*$_2$ (dash dot dot line) and T$_c$=T$_{3on}$. The zones 1 and 2 are due to the creep

and/or to the Flow, whereas the zone 3 is due to the TAFF and/or to the Flux Flow.

**Fig. 3:** *a*) Comparison between the temperature dependences of $\chi_1^{''}$ and $\chi_3^{'}$ in absence of DC magnetic field at $\nu$=1607 Hz, for $h_{AC}$=1 Oe, and *b*) for $h_{AC}$=16 Oe, showing the presence of the "zone 2" which increases for increasing AC field amplitudes. The vertical dashed lines mark the temperature $T_p$ of the peak in the $\chi_1^{''}(T)$ curves, and the dotted lines correspond to the temperature below which the $\chi_3^{'}(T)$ curves assume negative values. The data are reported in arbitrary units in order to compare the first and the third harmonics in the same plot.

**Fig. 4:** *a*) Comparison between the $\chi_1^{''}(T)$ and $\chi_3^{'}(T)$ at $H_{DC}$=0 and $\nu$=107 Hz, for $h_{AC}$=4 Oe, and *b*) for $h_{AC}$=32 Oe. The appearance of the "zone 2" for $h_{AC}$>8 Oe, which increases and melts with the "zone 3" as $h_{AC}$ is increased, indicates that at high enough AC magnetic fields it is not possible to determine a temperature region in which the AC magnetic response is governed by hysteretic phenomena. The vertical dashed lines mark the temperature $T_p$ for each measurement. The data are reported in arbitrary units in order to compare the first and the third harmonics in the same plot.

**Fig. 5**: Typical temperature dependences of $\chi_1''$ and $\chi_3'$ obtained with $H_{DC} \gg h_{AC}$, summarizing the analysis discussed in the text for $H_{DC} \neq 0$ (see also Fig.2). The "zone 3" previously found in Fig. 2 is absent here. In the "zone 1" (vertical lines filled) and in the "zone 2" (horizontal lines filled) the creep contribution can be detectable.

**Fig. 6: *a)*** $\chi_3'(T)$ measured at different $H_{DC}$ and at fixed fixed $\nu$ and $h_{AC}$: for $T > T_p$ the "zone 3" does not appear if $H_{DC} \neq 0$ (see the inset in this Figure); **Inset of Fig. 6*a*)**: Magnification of the region enclosed in the dashed rectangle and corresponding to the curve at $H_{DC}=0$ and at $H_{DC} \neq 0$

**Fig. 6: *b)*** $\chi_3''(T)$ curves corresponding to the $\chi_3'(T)$ curves in Fig. 6*a*): the $\chi_3''(T)$ curve exhibits a negative minimum close to $T_{3on}(\chi_3'')$ only if $H_{DC} \neq 0$ (see the inset of Fig. 6.b)); **Inset of Fig. 6*b*)**: Magnification of the region enclosed in the dashed rectangle of Fig. 6b), and corresponding to the curve at $H_{DC}=0$ and at $H_{DC} \neq 0$

**Fig. 7**: Plot of the $\chi_3'(T)$ measured at different $h_{AC}$, and at fixed $\nu$ and $H_{DC} \neq 0$. The vertical lines indicate the temperatures $T_p$ extracted from the corresponding $\chi_1''(T)$ curves: note the difference between the cases in which the "zone 2" exists ($h_{AC}$=1 Oe and $h_{AC}$=2 Oe) and the ones in which it is absent ($h_{AC}$=4 Oe and $h_{AC}$=8 Oe); **Inset of Fig. 7**: The behavior of the $\chi_3'(T)$ for 2 different frequencies and $H_{DC} \gg h_{AC}$.

**Fig. 8**: Experimental evidence of the $\Delta T_{on}$ indicating the region where the TAFF regime dominates the AC magnetic response, obtained by comparing the $\chi_1''(T)$ and $\chi_3''(T)$ with $H_{DC} \gg h_{AC}$; **Inset of Fig. 8**: Measurements of $\chi_1''(T)$ and $\chi_3''(T)$ with the same $h_{AC}$ and $\nu$ as in Fig.8, but with $H_{DC}$=0. In this case, no evidence of $\Delta T_{on} \neq 0$ can be found.

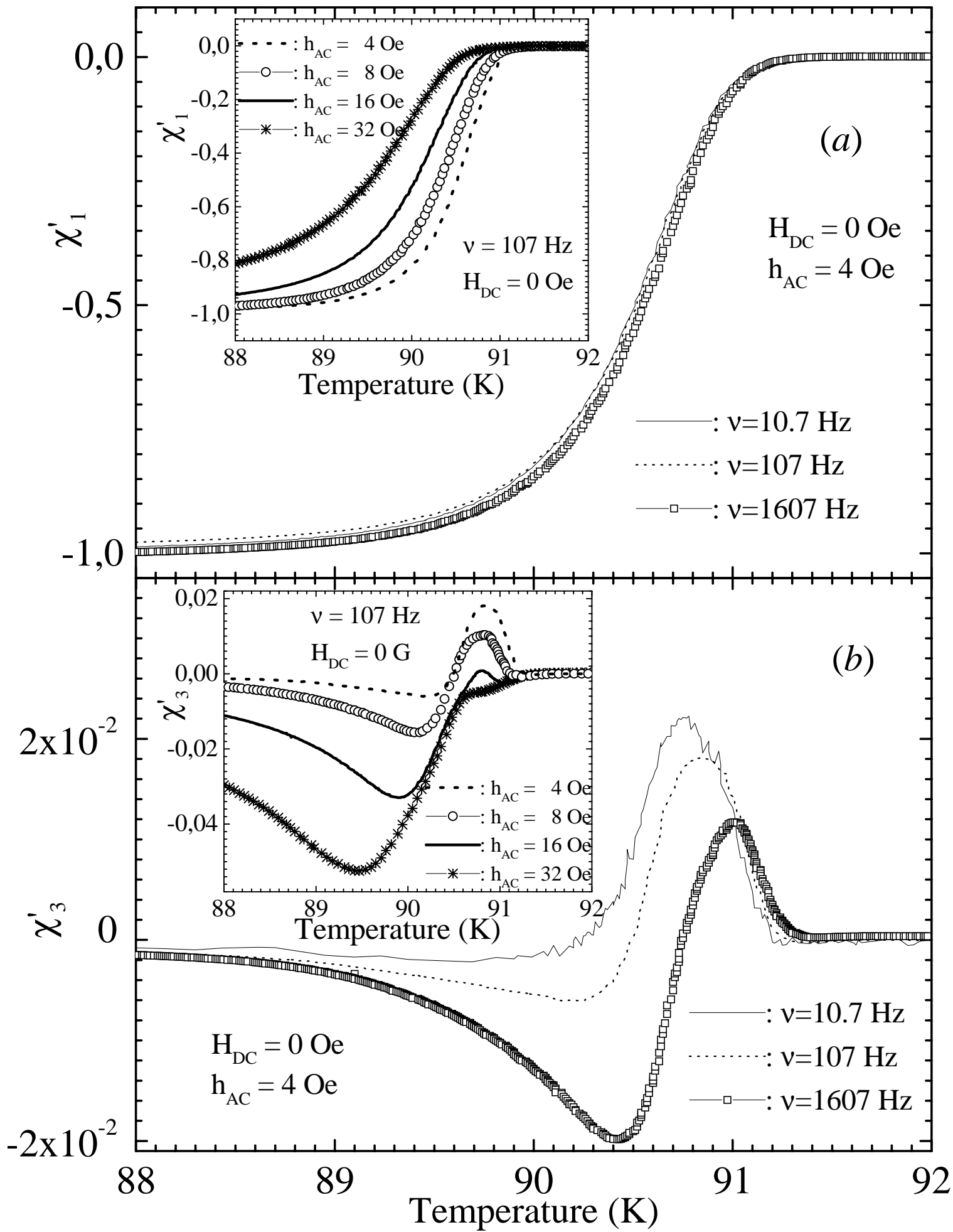

Fig.1, M. Polichetti et al.

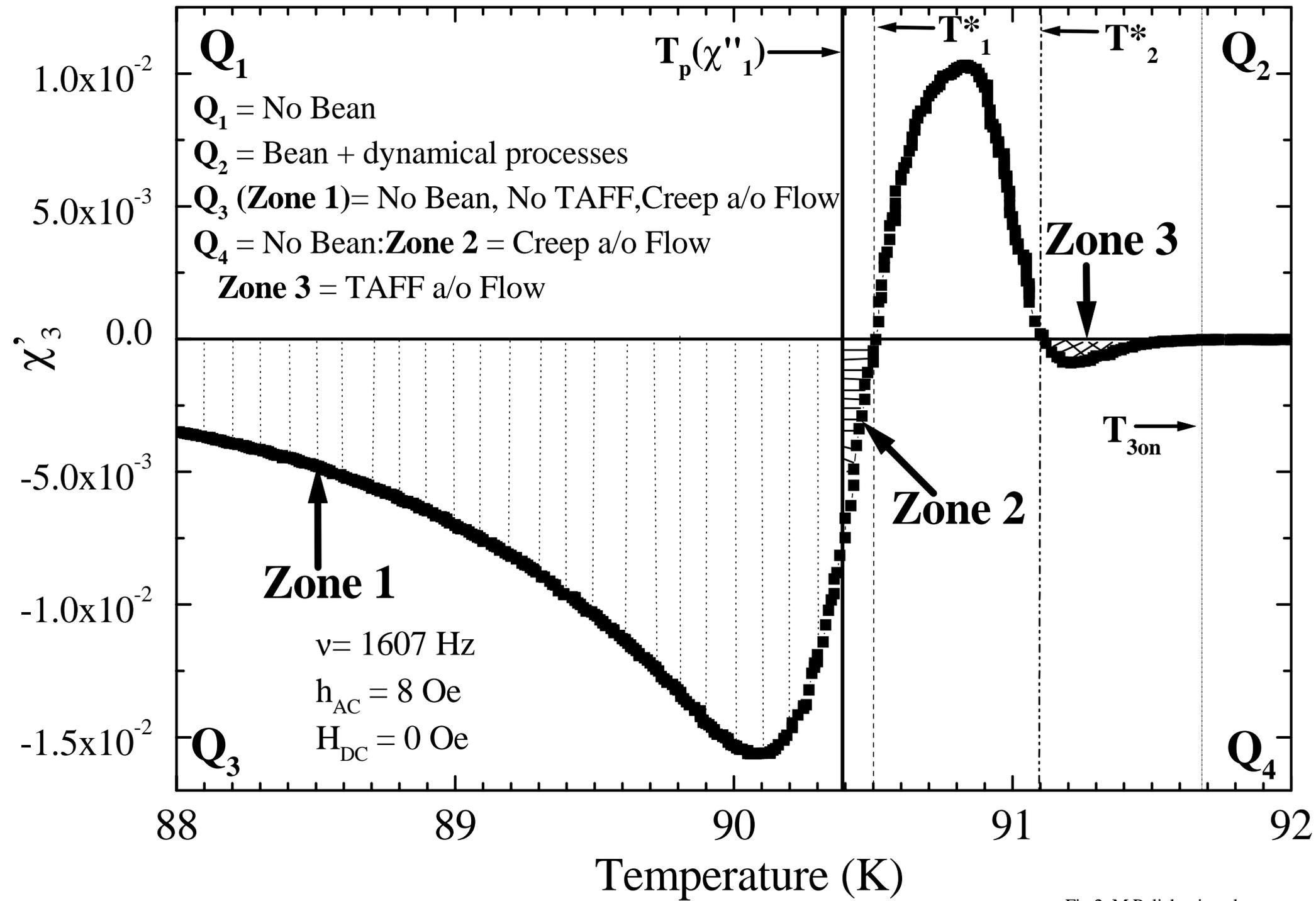

Fig.2, M.Polichetti et al.

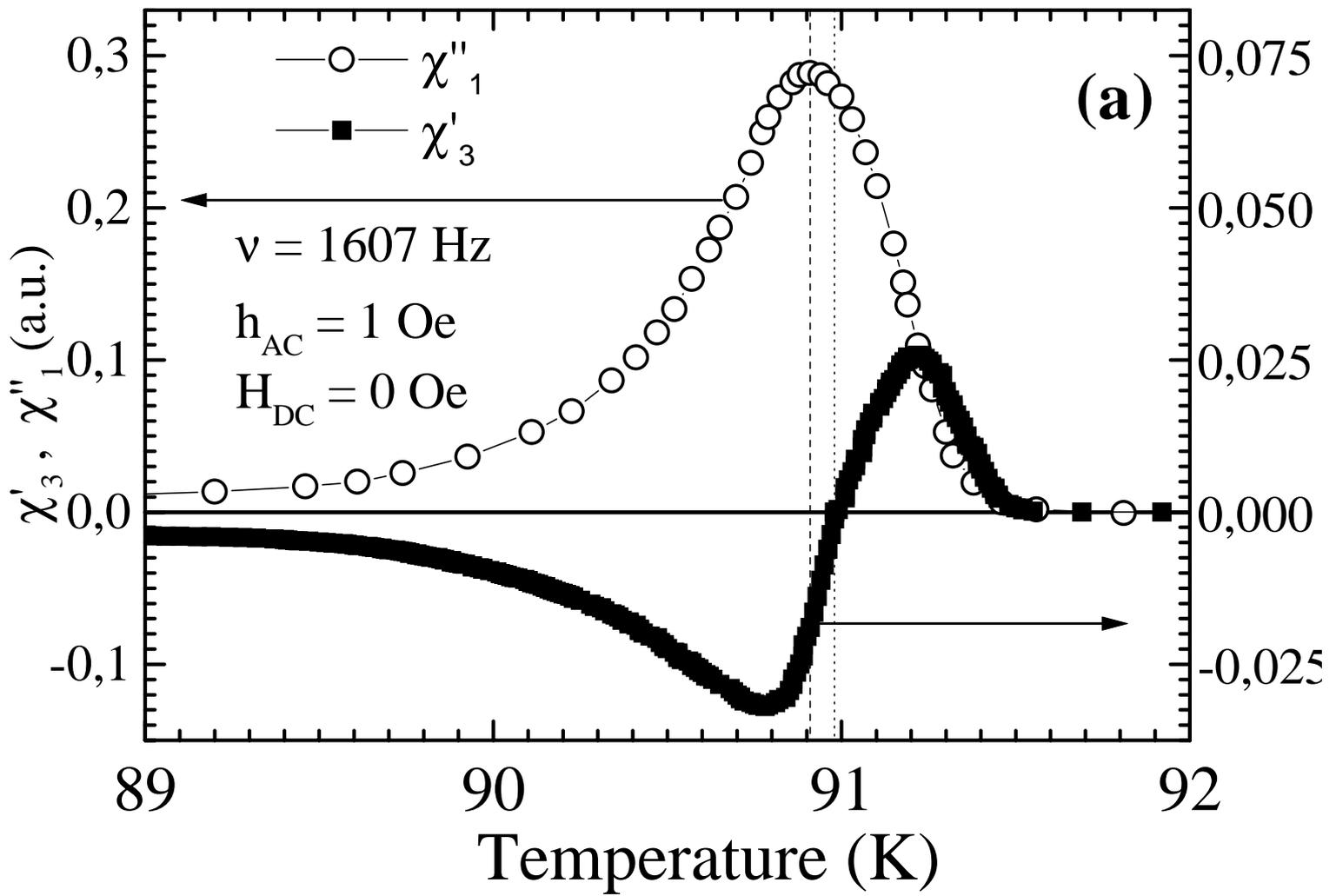

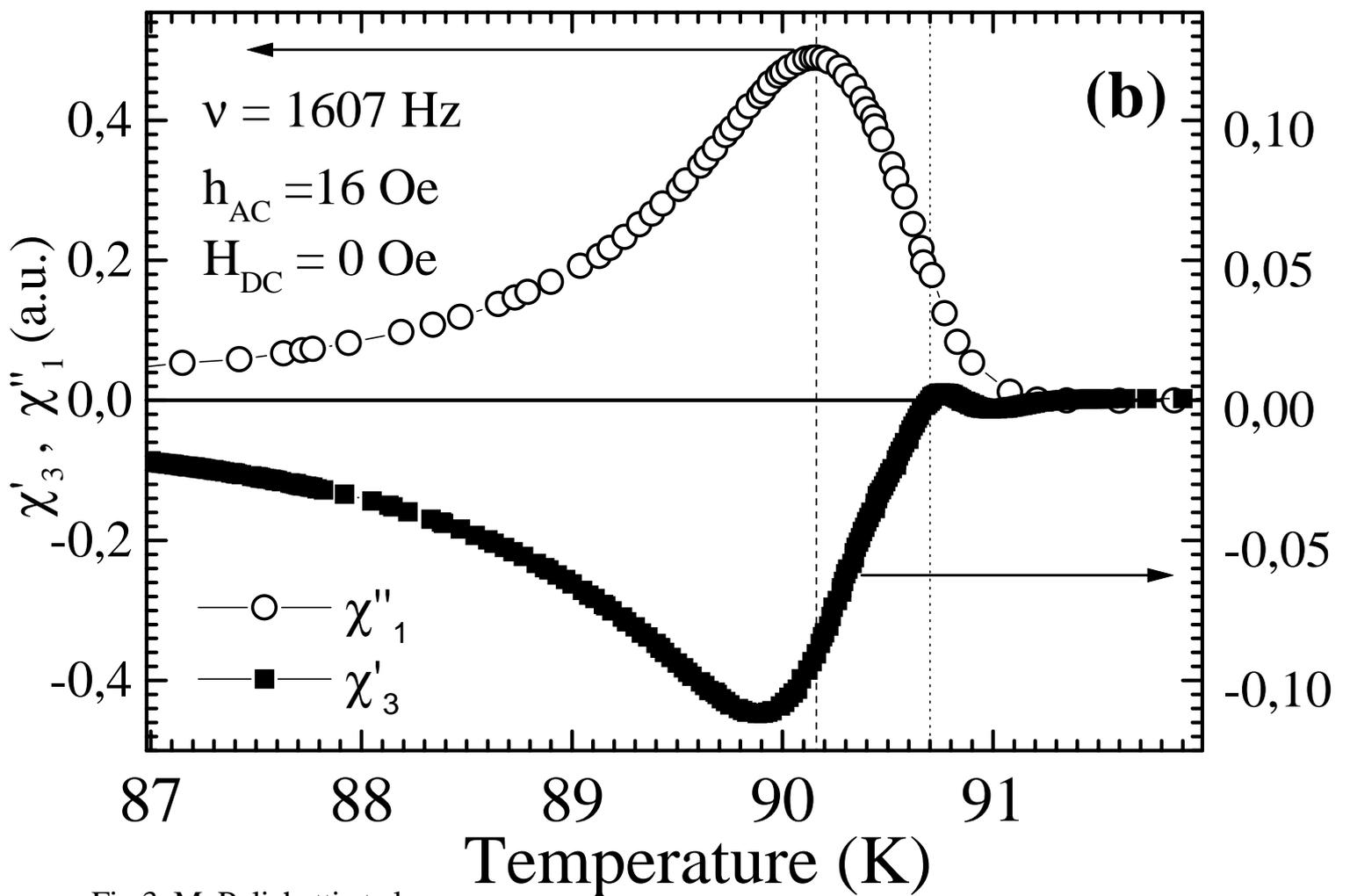

Fig.3, M. Polichetti et al.

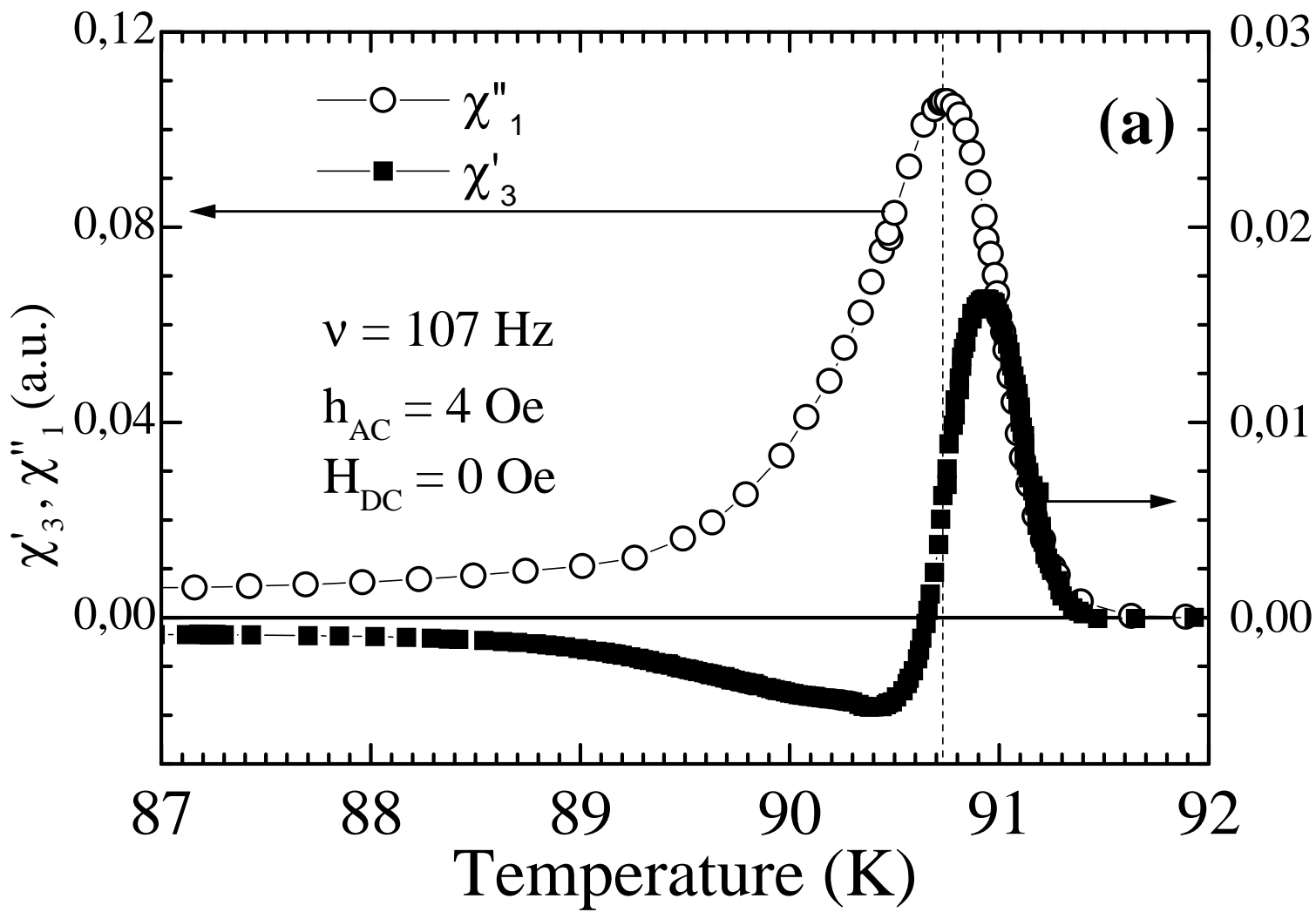
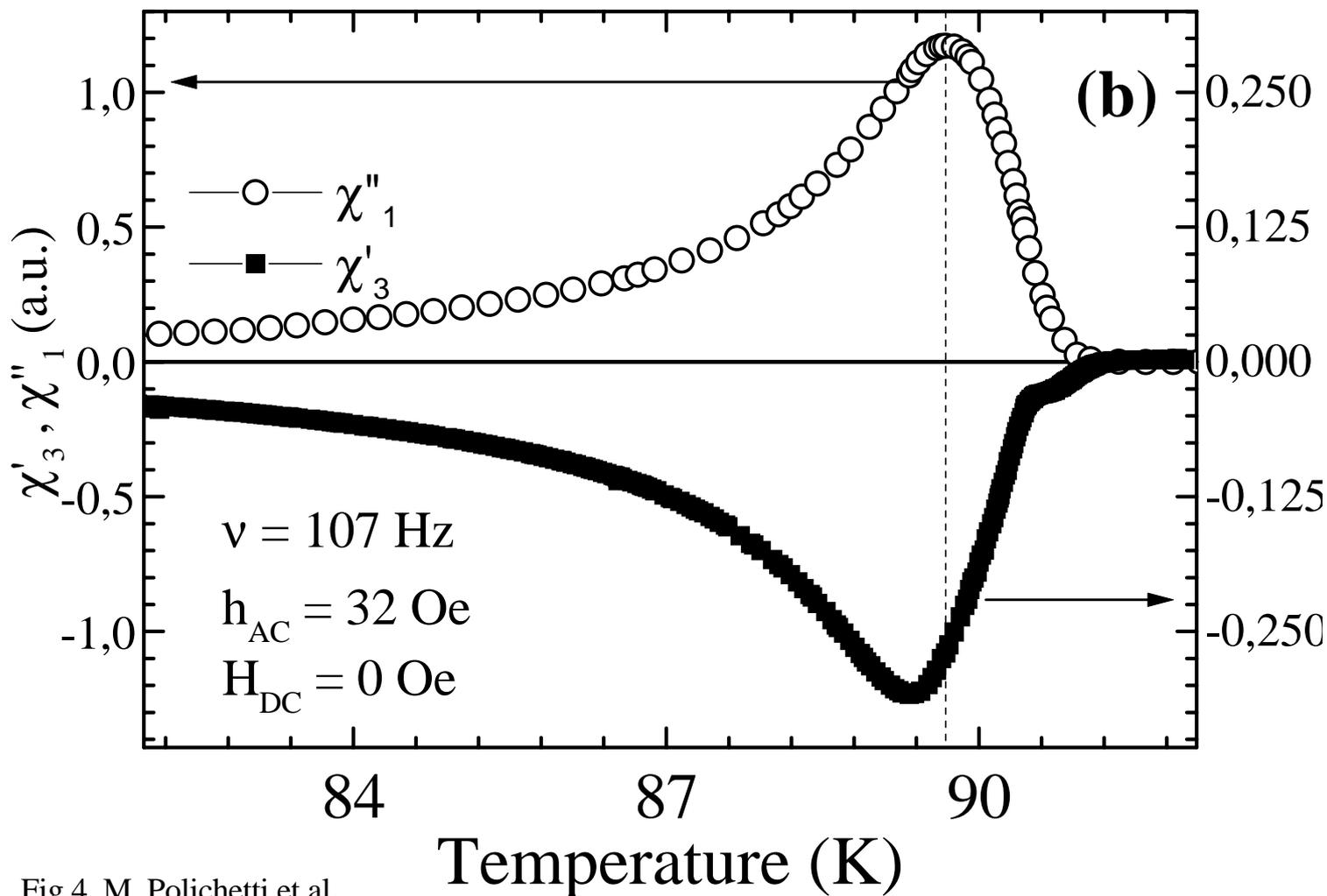
Fig.4, M. Polichetti et al.

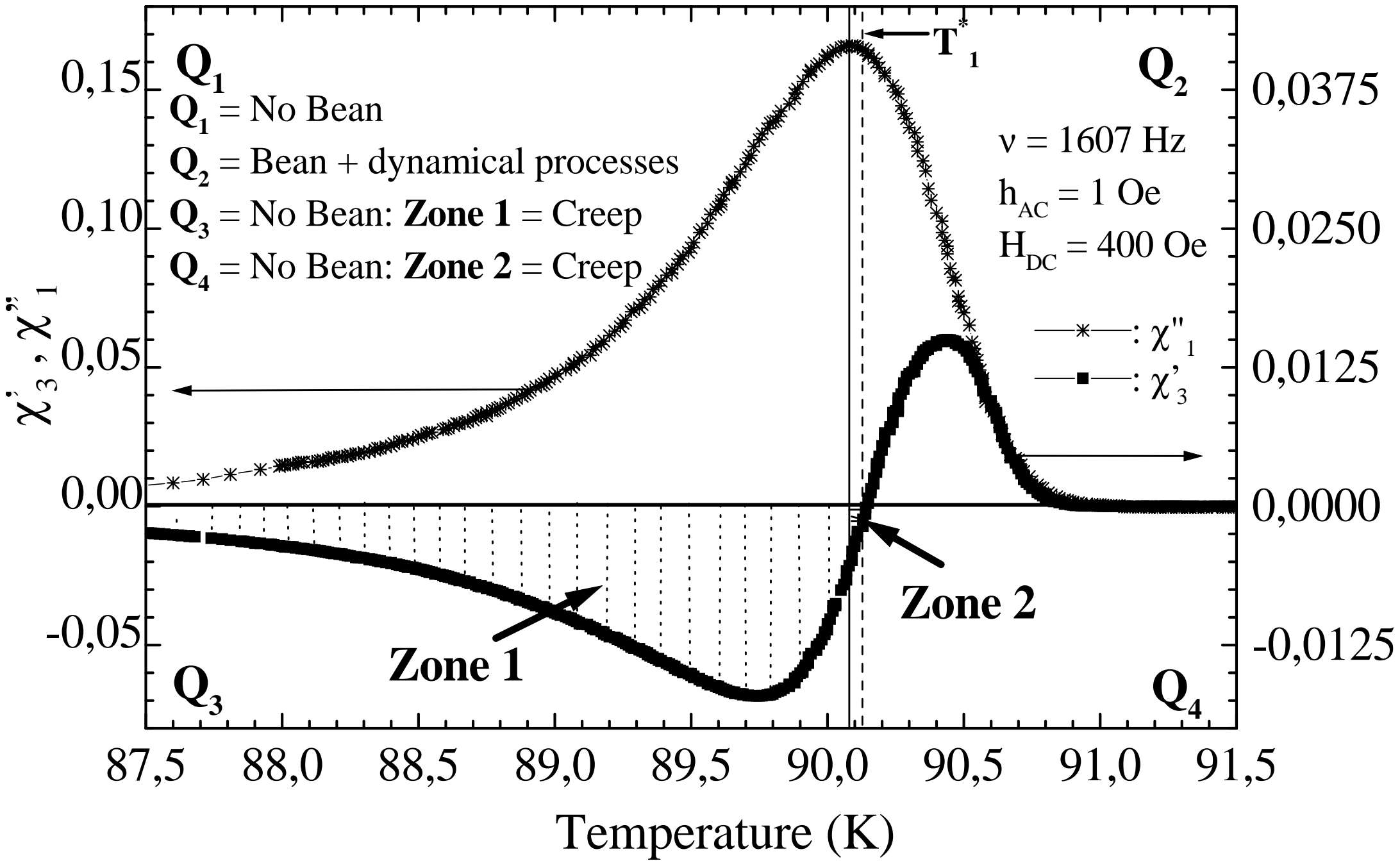

Fig.5, M. Polichetti et al.

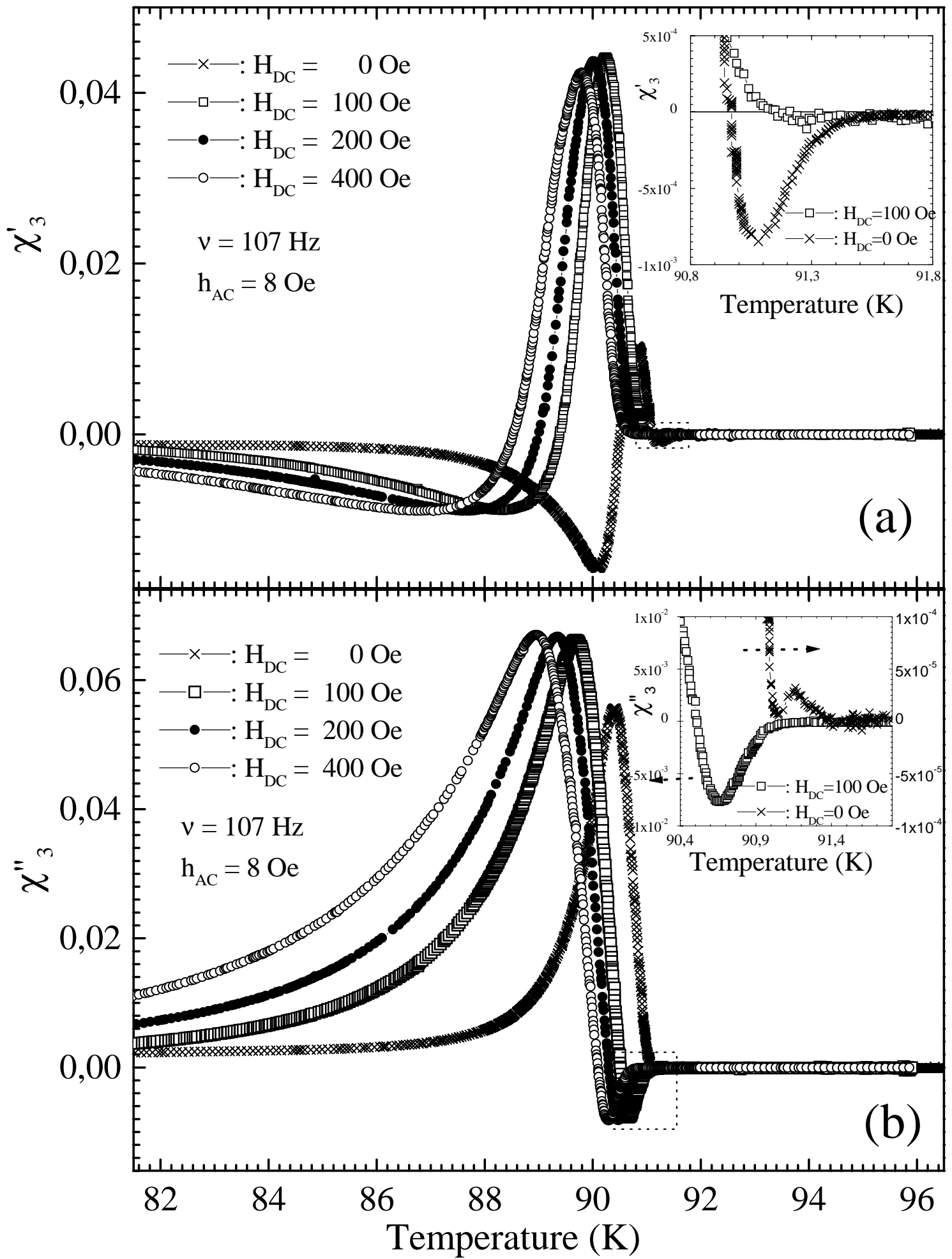

Fig. 6 M. Polichetti et al.

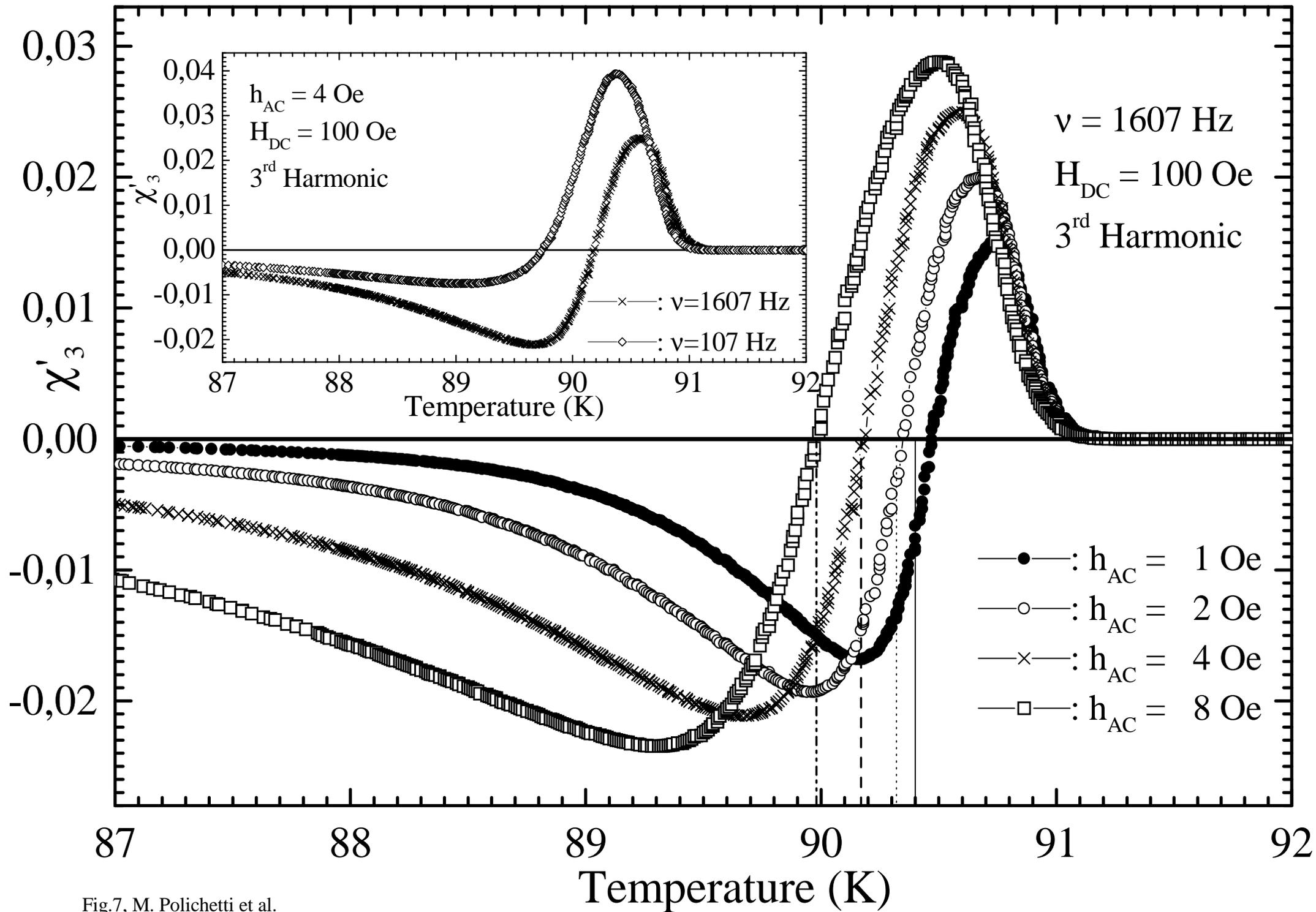

Fig.7, M. Polichetti et al.

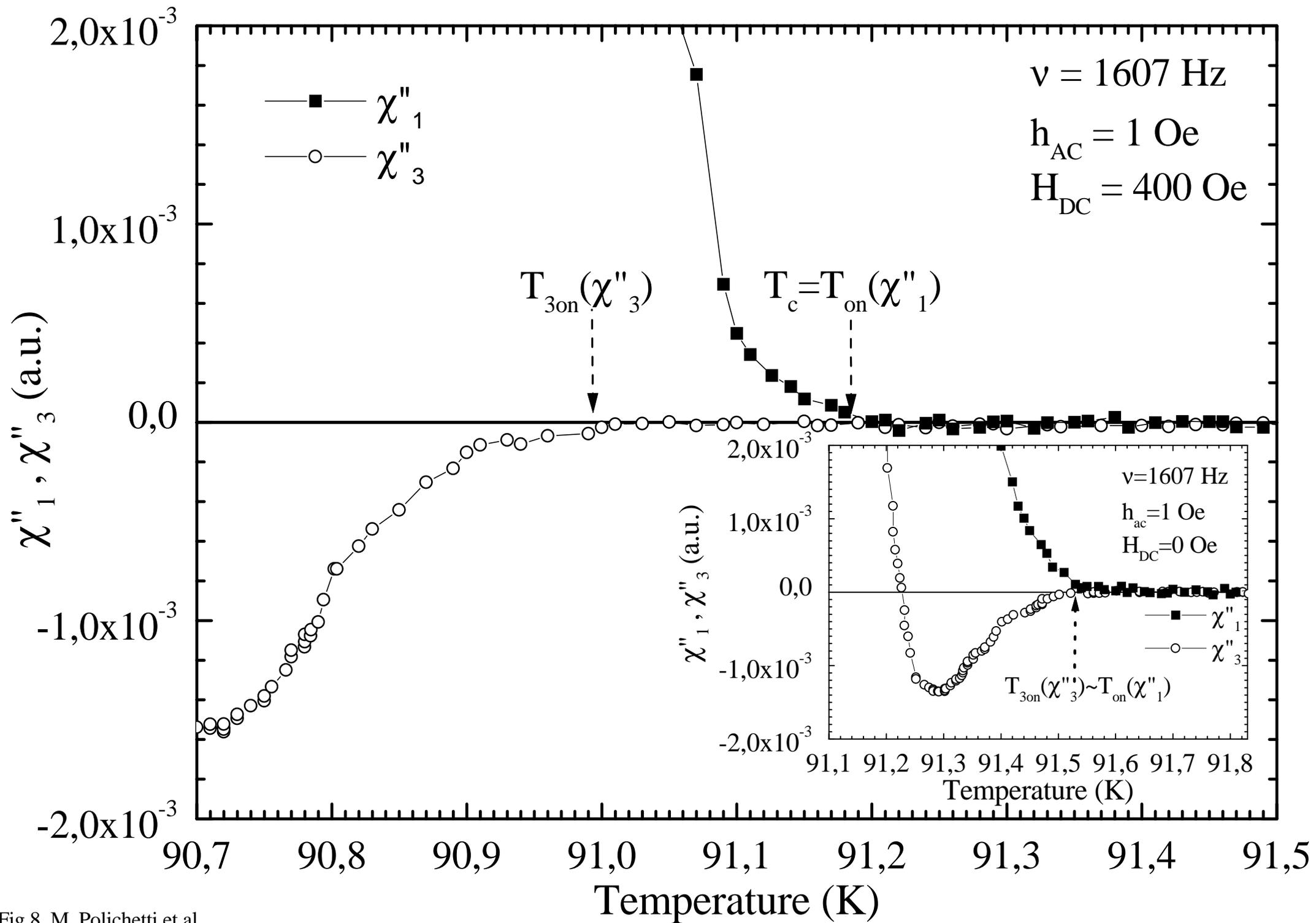

Fig.8, M. Polichetti et al.